\documentclass[11pt,letterpaper]{article}

\usepackage[left=1in,right=1in,top=1in,bottom=0.9in]{geometry}

\usepackage{graphicx}
\usepackage[table,xcdraw]{xcolor}
\usepackage{wrapfig,booktabs}
\usepackage{setspace}
\usepackage{tikz}

\usetikzlibrary{decorations.shapes,shapes.geometric,shadows,arrows,automata,positioning,calendar,mindmap,backgrounds,scopes,chains,er,3d,matrix,fit}

\usepackage[utf8]{inputenc}
\usepackage{amsmath}
\usepackage{amsfonts,times}
\usepackage{natbib}

\usepackage{lmodern}
\usepackage[most]{tcolorbox}

\usepackage{amssymb}
\usepackage{url}

\usepackage[compact]{titlesec}
    \titlespacing{\section}{0pt}{3ex}{2ex}
    \titlespacing{\subsection}{0pt}{2.0ex}{1.0ex}
    \titlespacing{\subsubsection}{0pt}{1.0ex}{0.5ex}

\usepackage{array}
\newcolumntype{L}{>{\centering\arraybackslash}m{3cm}}

\begin{document}

\setstretch{1.08}
\sloppy

\pagenumbering{gobble}
%\hfill \null
%\begin{center}
% {\bf \Large Supply Chain Cognizant Framework to Assist Procurement, Deployment, and Upgrade of IoT Systems for Security and Resilience of Critical Infrastructure }
% {\bf \Large Supply Chain Risk Assessment of IoT Infrastructure for Resilient Procurement and Deployment Decisions}
%\linespread{1.01}

%{\bf \large  U.S.-Ireland R\&D Partnership: SaTC: CORE: Small: Enhancing Resilience and Safety in Industrial Control Systems through Integrated IT and OT Modeling and Analysis \par}
% \end{center}

%\vspace{0.5em}

%\begin{figure}[b!]
%    \centering
%    \vspace{-0.6cm}
%    \includegraphics[width=\columnwidth]{summary_fig.pdf}
%    \vspace{-1.3in}\caption{Overview of the stages and components of the proposed research.\vspace{-0.2in}}
%    \label{fig:summary}
%\end{figure}
%\newpage

%\setcounter{page}{1}
%\def\thepage{C -- \arabic{page}}

\title{Cooperative and Noncooperative Paradigms for Game-Theoretic Control of Socio-Technical Systems\thanks{This paper is prepared for the tutorial at the IFAC World Congress Tutorial 2026, Busan, Korea.}}

\author{
Tamer Ba\c{s}ar\thanks{Department of Electrical and Computer Engineering, University of Illinois Urbana-Champaign, Urbana, IL, USA (e-mail: basar1@illinois.edu).}
\and
Tomohisa Hayakawa\thanks{Department of Electrical and Electronic Engineering, Tokyo Institute of Technology, Tokyo, Japan (e-mail: hayakawa@ee.e.titech.ac.jp).}
\and
Hideaki Ishii\thanks{Department of Information Physics and Computing, The University of Tokyo, Tokyo, Japan (e-mail: ishii@ipc.i.u-tokyo.ac.jp).}
\and
Quanyan Zhu\thanks{Department of Electrical and Computer Engineering, New York University, NY, USA (e-mail: qz494@nyu.edu).}
}

\date{}

\maketitle

\begin{abstract}
This tutorial presents cooperative and noncooperative game-theoretic frameworks for modeling, learning, and control in socio-technical systems, where human behavior, incentives, institutions, and social interactions are coupled with cyber-physical and networked infrastructures. The paper reviews strategic, dynamic, cooperative, matching, learning, and feedback-control approaches for analyzing how local decision-making, adaptation, and strategic interactions shape collective system outcomes. The tutorial further develops feedback-learning and incentive-design perspectives that connect equilibrium analysis with adaptation, distributed control, and mechanism design under information and coordination constraints. We also examine resilience and security challenges arising from adversarial behavior, misinformation, disruptions, and cascading failures in interconnected socio-technical networks. Finally, we discuss emerging research directions at the intersection of game theory, control, learning, and network science for resilient and adaptive socio-technical systems.
\end{abstract}


\vspace{1em}
%===============================================================================

\section{Introduction}

\begin{figure}[h]
    \centering
    \includegraphics[width=0.93\columnwidth]{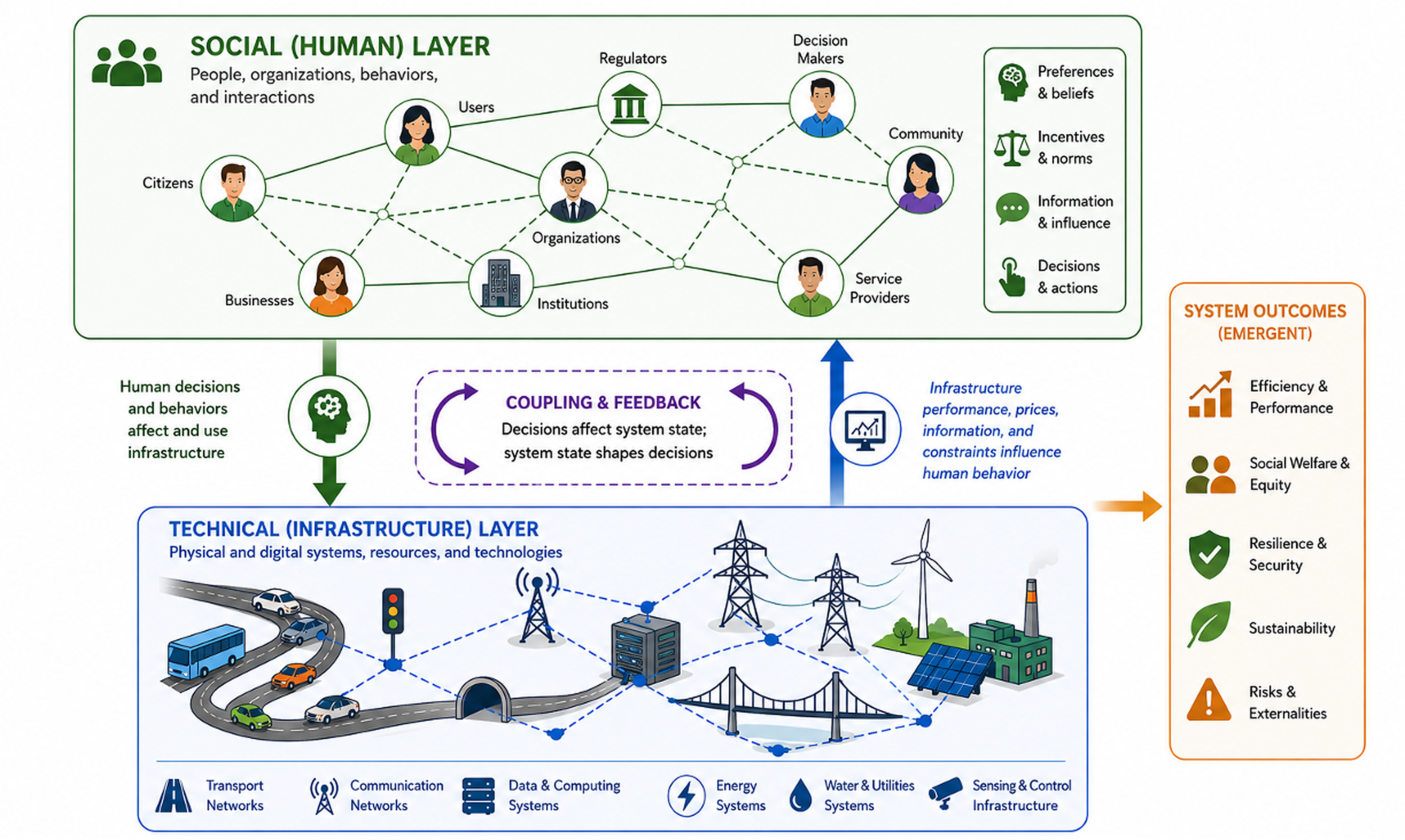}
    \caption{
    Socio-technical systems consist of coupled social and technical layers.
    Human behaviors, incentives, and institutions interact with infrastructures, cyber-physical systems, and digital platforms through continuous feedback and adaptation, leading to emergent system-level outcomes such as efficiency, resilience, fairness, and sustainability.
    }
    \label{fig:socio_technical_system}
\end{figure}

Socio-technical systems couple human behavior, incentives, institutions, and social influence with technological infrastructures, cyber-physical systems, communication networks, transportation systems, energy systems, and digital platforms \citep{huang2023cognitive,zhu2025revisiting,zhu2013networked}. As illustrated in Fig.~\ref{fig:socio_technical_system}, these layers cannot be studied independently: technology shapes human behavior, while human actions and collective responses alter the operation and performance of the technical system. Congestion pricing provides a representative example. Transportation authorities may introduce tolls to reduce congestion and improve traffic efficiency; drivers respond by adjusting routes, departure times, or travel behavior; and these adaptations reshape traffic conditions and the effectiveness of the pricing policy itself.

The tight coupling between social and technical components makes these systems difficult to understand, predict, and control. Technical systems are often modeled using physical laws, dynamical systems, optimization, and control theory, whereas human and social systems are commonly studied through economics, psychology, and behavioral sciences. When these layers interact, emergent behaviors arise that cannot be understood from either perspective in isolation. A classical example is the Braess paradox \citep{braess1968}, where adding infrastructure to a transportation network can worsen congestion because individually rational routing decisions alter the global traffic equilibrium. The design objective is therefore not merely to optimize isolated components, but to achieve system-level properties such as efficiency, resilience, fairness, robustness, safety, and social welfare while accounting for incentives, adaptation, feedback, and strategic behavior.

\subsection{Game-Theoretic Paradigms}

Game theory provides a natural framework for studying socio-technical systems because it explicitly models interactions among decision-making entities. First, it captures strategic human behavior while connecting naturally to engineering disciplines such as control systems \citep{bacsar1998dynamic}, distributed optimization, and network dynamics \citep{jackson2008social,li2022confluence,chen2019game}. This establishes a unified mathematical language for social and technical components.

Second, game theory explains how local interactions among decentralized agents generate collective outcomes such as congestion, cooperation, coordination, instability, resilience, and social influence. It also provides analytical and computational tools for mechanism design, learning dynamics, incentive engineering, distributed algorithms, and scalable networked or large-population models. Finally, socio-technical systems typically involve both competitive and cooperative interactions. Individuals and organizations may compete for resources, utility, or influence while simultaneously coordinating or collaborating to achieve collective objectives, making game theory particularly suitable for modeling mixed interaction structures.

\subsection{Tutorial Overview}

In this tutorial, we present game-theoretic methods as a foundational framework for understanding, analyzing, and designing socio-technical systems. The emphasis is on how local interactions among decision-making entities collectively shape global system outcomes.

In Section~\ref{sec:tamer} and Section~\ref{sec:quanyan}, we introduce methods from both noncooperative and cooperative game theory, respectively. Socio-technical systems are rarely purely competitive or purely cooperative; instead, they involve intertwined processes of competition, coordination, collaboration, bargaining, and coalition formation. For example, users in transportation networks may compete for limited resources while benefiting from coordination mechanisms that improve overall efficiency \citep{carlier2008optimal}. Similarly, participants in cybersecurity ecosystems may pursue conflicting objectives while sharing incentives for collective resilience and information exchange \citep{chen2019interdependent}.

The tutorial introduces fundamental modeling frameworks and solution concepts for these settings. Noncooperative frameworks in Section~\ref{sec:tamer} include strategic-form games, dynamic games, Stackelberg games, and network games, whereas cooperative frameworks in Section~\ref{sec:quanyan} include coalition formation games, bargaining models, matching mechanisms, and cost-sharing formulations. Together, they characterize equilibria, incentives, cooperation structures, collective behaviors, efficiency, fairness, resilience, social welfare, and systemic risk.

Next, Section~\ref{sec:control} introduces a feedback control paradigm based on game-theoretic models, emphasizing learning, adaptation, estimation, and coordination through technical interventions such as resource allocation and routing. Section~\ref{sec:tomohisa} then focuses on incentive mechanisms, including hierarchical and budget-constrained designs.

In Section~\ref{sec:hideaki}, we examine resilience aspects of socio-technical systems, including adversarial behaviors, anomalies, disruptions, cyberattacks, misinformation propagation, and strategic manipulation. Section~\ref{sec:conclusion} concludes the tutorial and presents several future research directions.

\section{Noncooperative Game-Theoretic Methods}\label{sec:tamer}

Noncooperative game theory models situations in which each decision-making
agent optimizes its own objective while anticipating the decisions of others.
This viewpoint is central for socio-technical systems because congestion,
security risk, market outcomes, platform behavior, and infrastructure use are
often shaped by many decentralized choices rather than by a single centralized
optimizer. Classical game theory provides the equilibrium language for these
interactions \citep{vonNeumann1944,nash1950,nash1951,fudenberg1991}, while
dynamic games connect this language to control, information, and temporal
adaptation \citep{BZ18,bacsar1998dynamic}. This section introduces the noncooperative modeling primitives
used throughout the tutorial. The notation developed here is used again in
the feedback-control and incentive-design sections,
Sections~\ref{sec:control} and~\ref{sec:tomohisa}.

\subsection{Strategic-Form Games}

Consider a set of agents $\mathcal{N}=\{1,\dots,N\}$. Agent $i$ chooses an
action or strategy $x_i\in\mathcal{X}_i$, and the joint action profile is
$x=(x_i,x_{-i})$, where $x_{-i}$ collects the actions of all agents except
$i$. The payoff of agent $i$ is denoted by $J_i(x_i,x_{-i};c)$, where the
optional argument $c$ represents an external parameter, coordination signal,
institutional rule, network condition, or environmental state that shapes the
game. A Nash equilibrium is an action profile $x^\ast$ such that
\[
J_i(x_i^\ast,x_{-i}^\ast;c)\geq J_i(x_i,x_{-i}^\ast;c),
\qquad
\forall x_i\in\mathcal{X}_i,\ i\in\mathcal{N}.
\]
Thus, no agent can improve its own payoff by changing its action unilaterally.
This unilateral optimality condition is the basic noncooperative analogue of
collective agreement: stability is defined not by global optimality, but by
the absence of profitable individual deviations.

This distinction is especially important in socio-technical systems. A Nash
equilibrium may be inefficient, unfair, fragile, or socially undesirable even
when every agent is acting rationally with respect to its own objective. In
transportation and communication networks, for example, selfish routing can
increase congestion relative to coordinated routing; in digital platforms,
locally optimal user or algorithmic behavior can produce polarization,
manipulation, or degraded information quality; and in cybersecurity, defensive
investments can be strategically coupled across organizations. These issues
motivate performance notions such as social welfare, price of anarchy,
incentive compatibility, and robustness \citep{roughgarden2005}.
They also motivate the control-oriented question studied later: how should
signals, incentives, and constraints be designed so that decentralized
equilibria align with system-level goals?

\subsection{Dynamic and Stochastic Games}

Many socio-technical systems are not one-shot interactions. Agents observe,
learn, adapt, and repeatedly respond to changing conditions. To distinguish
the physical or informational environment from the agents' decisions, let
$z_t$ denote the system state at time $t$, and let
$x_t=(x_{1,t},\dots,x_{N,t})$ denote the joint action profile. A general
controlled evolution can be written as
$z_{t+1}=f_t(z_t,x_t,w_t)$, where $w_t$ captures uncertainty, disturbances,
or unmodeled behavior. Agent $i$ evaluates a stream of payoffs such as
$\mathbb{E}[\sum_{t\geq0}\beta_i^t J_i(z_t,x_{i,t},x_{-i,t})]$, with
discount factor $\beta_i\in(0,1)$ when an infinite-horizon discounted model is
appropriate.

A strategy in a dynamic game is a policy rather than a single action. We write
$I_{i,t}$ for the information available to agent $i$ at time $t$, and
$\gamma_{i,t}:I_{i,t}\mapsto\mathcal{X}_i$ for a policy that maps information
to an action. The information set may contain local measurements, neighboring
actions, prices, recommendations, histories, beliefs, or public signals. The
choice of information structure is not a technical detail; it determines what
forms of strategic reasoning and adaptation are feasible. Open-loop policies
depend mainly on initial information, feedback policies respond to observed
states or signals, and partially observed policies must act under incomplete
and noisy information \citep{akyol2016information}.

An equilibrium of a dynamic game is a profile of policies
$\gamma^\ast=(\gamma_1^\ast,\dots,\gamma_N^\ast)$ such that no agent can
improve its expected payoff by replacing its own policy while the other
policies remain fixed. Dynamic equilibria therefore extend Nash's unilateral
optimality idea from actions to information-dependent decision rules. They are
well suited for networked control, security, energy, transportation, and
communication problems where actions affect both current payoffs and future
states \citep{bacsar1998dynamic}. They also clarify why learning and control cannot be separated: when agents
adapt their policies over time, the equilibrium concept describes a possible
steady outcome, while the learning dynamics determine whether that outcome can
actually emerge.

\subsection{Stackelberg and Design Perspectives}

Socio-technical systems often include a designer, regulator, platform, or
infrastructure operator that moves before the agents. This creates a
hierarchical game. The leader selects a signal, incentive, pricing rule,
information policy, or constraint $c$, and the followers then play a
noncooperative game parameterized by $c$. If $x^\ast(c)$ denotes a follower
equilibrium, the leader's design problem can be viewed as choosing $c$ to
optimize $J_0(c,x^\ast(c))$, where $J_0$ is the system-level objective.

This Stackelberg viewpoint is useful because many interventions in
socio-technical systems are indirect. A transportation authority typically
sets tolls or recommendations rather than directly choosing every driver's
route. A grid operator sends prices or demand-response incentives rather than
controlling every appliance. A cyber defender hardens assets, allocates
monitoring resources, or reveals information while anticipating adversarial
adaptation. Similar leader-follower models arise in communications, smart
grids, security, and transportation
\citep{yang2007stackelberg,chen2017stackelberggameapproachtwo,SBB24}.

The design problem is delicate because follower equilibria may be multiple,
unstable, or hard to compute. Optimistic and pessimistic Stackelberg models
handle equilibrium multiplicity by assuming, respectively, favorable or
adverse follower equilibrium selection. In applications, the relevant
selection is often induced by learning dynamics, institutions, conventions, or
the timing of information. Thus, Stackelberg design naturally connects the
static language of equilibrium with the dynamic language of feedback,
adaptation, and control.

\subsection{Network and Population Games}

Interactions in socio-technical systems are rarely all-to-all. Agents are
embedded in social, physical, cyber, economic, or informational networks.
Let $\mathcal{G}_{i,t}$ denote the neighbors that directly influence agent
$i$ at time $t$. A network game allows $J_i$ to depend on the local profile
$(x_i,x_{\mathcal{G}_{i,t}})$, on aggregate quantities such as congestion or
load, and on external coordination signals. This local dependence captures
peer effects, routing externalities, epidemic risk, market spillovers,
infrastructure interdependence, and strategic complementarities or
substitutes \citep{jackson2008social,goyal2023networks,li2022role}.

When the number of agents is large, population and mean-field models replace
fine-grained strategic coupling with aggregate state variables such as an
empirical distribution $m_t$. Agent $i$ then responds to its own state, local
information, and the population aggregate, while the aggregate evolves from
the collective response of all agents. This approximation is especially useful
for large infrastructures, cyber-physical systems, epidemics, and massive
online platforms, where tracking every bilateral interaction is infeasible
\citep{caines2021mean,basar2020recent}. Related
network-equilibrium models also appear in congestion and routing problems,
including Wardrop equilibria, price-of-information effects, and robust or
adversarial variants \citep{carlier2008optimal,zhu2010price}.

The noncooperative models in this section provide the baseline for the rest
of the tutorial: they explain how decentralized incentives, information, and
time shape equilibrium behavior. The next section complements this view by
asking when agents can form coalitions, share value, bargain, match, and
cooperate.

\section{Cooperative Game-Theoretic Methods}\label{sec:quanyan}

Cooperative game theory complements the noncooperative perspective of
Section~\ref{sec:tamer} by asking
when agents can form coalitions, how much value cooperation creates, and how
that value should be allocated among participants \citep{maschler2013game}. In
noncooperative games, stability is expressed through unilateral deviations. In
cooperative games, stability is instead tied to collective deviations: a group
of agents may leave the proposed arrangement if it can generate and distribute
more value on its own.

This perspective is useful for socio-technical systems because many design
problems involve both competition and cooperation. Robots may coordinate to
complete missions that are impossible individually; communication providers may
share infrastructure or spectrum; organizations may share threat intelligence;
and participants in federated learning may jointly train models while still
needing credible reward allocation rules \citep{kairouz2021advances,zhang2021survey}.
Similar cooperation questions arise in collaborative intrusion detection,
networked security, resource matching, and cyber-physical systems where agents
must decide when to share data, rules, trust, or operational resources
\citep{zhu2010distributedsequentialalgorithmco,fung2016facidtrustbasedcollaborative,zhu2011gametheoreticapproachknowledge,zhu2009gametheoreticalapproachincentive}.
Cooperative game theory provides a language for coalition value, stability,
fairness, bargaining, matching, and cost sharing.

In socio-technical systems, cooperative games are useful whenever system
performance depends on joint action rather than isolated optimization. In
transportation, they can model ride sharing, route pooling, fleet coordination,
and cost allocation for shared infrastructure. In energy systems, they support
coalitions of distributed energy resources, microgrids, storage devices, and
flexible loads that jointly provide reliability, resilience, or market
services. In cybersecurity, they describe information sharing, collaborative
intrusion detection, cyber-insurance pools, and coordinated recovery after
attacks, contract-based security services, and cyber-insurance pools
\citep{chen2017security,zhang2019gametheoreticcyberinsurance}. In communication and computation networks, they capture spectrum
sharing, cloud-resource pooling, edge-computing coalitions, and federated
learning participation. Contract-theoretic and cooperative formulations in
resilient microgrids illustrate how such coalitions can align investment,
recovery, and operational incentives \citep{chen2022transactive,chen2016game}.
Across these examples, cooperative games help answer
three recurring design questions: which coalitions are valuable, which
allocations keep participants willing to cooperate, and which mechanisms make
cooperation stable under strategic incentives and operational constraints.

\subsection{Characteristic Function Form}

Using the agent set $\mathcal{N}$ introduced in Section~\ref{sec:tamer}, a
transferable-utility cooperative game is described by a characteristic function
$v:2^{\mathcal{N}}\rightarrow\mathbb{R}$, where $v(S)$ is the value that
coalition $S\subseteq\mathcal{N}$ can generate by cooperating. The value may
represent profit, throughput, resilience, mission success probability,
estimation quality, computational gain, or any other performance measure that
can be shared among coalition members.

The characteristic function captures how collective capability changes as
agents combine resources, information, and actions. A simple three-agent game
has agents indexed by $\{1,2,3\}$, $v(\emptyset)=0$,
$v(\{1\})=v(\{2\})=v(\{3\})=0$,
$v(\{1,2\})=v(\{1,3\})=v(\{2,3\})=1/2$, and
$v(\{1,2,3\})=1$. In this game, no agent creates value alone, pairwise
cooperation creates positive value, and full cooperation creates the largest
total value.

A game is superadditive when disjoint coalitions never lose value by merging:
$v(S\cup T)\geq v(S)+v(T)$ for all disjoint $S,T\subseteq\mathcal{N}$.
Superadditivity formalizes synergy. It appears when communication providers
reduce costs by sharing infrastructure, distributed sensors improve estimation
by pooling measurements, or cyber organizations improve detection by sharing
signals \citep{basar2011pricesanarchyinformationcooperat}. The cooperative surplus,
$v(\mathcal{N})-\sum_{i\in\mathcal{N}}v(\{i\})$, measures the additional value
created by cooperation beyond what agents can achieve separately.

\subsection{Allocations, Core, and Stability}

Once the grand coalition forms, its value must be distributed. We write
$r=(r_1,\dots,r_N)$ for an allocation, where $r_i$ is the payoff assigned to
agent $i$. An allocation is individually rational and efficient when
$r_i\geq v(\{i\})$ for every $i\in\mathcal{N}$ and
$\sum_{i\in\mathcal{N}}r_i=v(\mathcal{N})$. Individual rationality ensures
that no agent receives less than its stand-alone value, while efficiency
requires that the entire value of the grand coalition is allocated.

For the three-agent example above, the efficient individually rational
allocations are the simplex
\[
\left\{r\in\mathbb{R}^3:\ r_1+r_2+r_3=1,\ r_i\geq0\right\}.
\]
This set describes allocations that satisfy individual participation
constraints, but it does not yet guarantee that groups of agents want to remain
in the grand coalition.

The core strengthens individual rationality to coalitional rationality. An
allocation $r$ lies in the core when it is efficient and
$\sum_{i\in S}r_i\geq v(S)$ for every coalition $S\subseteq\mathcal{N}$. The
condition says that each coalition receives at least as much value, in total,
as it could produce by itself. If this condition fails for some $S$, then that
coalition can block the proposed allocation by leaving and redistributing
$v(S)$ among its members.

In the three-agent example, the core is obtained by adding the pairwise
constraints $r_1+r_2\geq1/2$, $r_1+r_3\geq1/2$, and
$r_2+r_3\geq1/2$ to the simplex. Allocations satisfying these inequalities are
stable against deviations by all one-agent and two-agent coalitions.

The core may be empty. The Bondareva-Shapley condition characterizes the
nonemptiness of the core through balancedness: a cooperative game has a
nonempty core exactly when coalitional demands are mutually compatible with
the total value of the grand coalition. This result links cooperative
stability to linear optimization and duality \citep{maschler2013game}. When
the core is empty, no allocation can satisfy all coalitional participation
constraints simultaneously, and the model points to an inherent instability in
the proposed cooperative arrangement.

Another stability-oriented concept is the nucleolus. For an allocation $r$,
the excess of coalition $S$ is $e(S,r)=v(S)-\sum_{i\in S}r_i$. A positive
excess means that the coalition is undercompensated relative to what it can
produce alone. The nucleolus chooses the efficient allocation that
lexicographically minimizes the ordered vector of coalition excesses. In this
sense, it balances the grievances of the most dissatisfied coalitions first,
then the next most dissatisfied coalitions, and so on. This makes it useful in
settings where robust fairness and worst-case dissatisfaction are central.
Related coalitional formulations for networked control and predictive control
show how coalition values can encode both performance and coordination costs
\citep{gomez2022coalitionalstochasticdifferentia,chanfreut2020noregretlearningcoalitional}.

\subsection{Shapley Value and Fairness}

The core focuses on stability, but it can contain many allocations and may not
select a unique or intuitively fair one. The Shapley value provides a
contribution-based fairness rule. For agent $i$, it is
\[
\phi_i(v)=
\sum_{S\subseteq\mathcal{N}\setminus\{i\}}
\frac{|S|!(N-|S|-1)!}{N!}
\left(v(S\cup\{i\})-v(S)\right).
\]
The term $v(S\cup\{i\})-v(S)$ is the marginal contribution of agent $i$ when
joining coalition $S$, and the coefficient averages this contribution over all
possible coalition formation orders.

The Shapley value is the unique allocation rule satisfying efficiency,
symmetry, the dummy-player property, and additivity. These axioms express that
all value is allocated, identical contributors are treated identically,
agents who add no marginal value receive no reward, and allocations are
consistent across additive combinations of games \citep{maschler2013game}.

This contribution-based view has become important in modern data-driven
systems. In explainable AI, Shapley values attribute prediction outcomes to
features. In federated learning, they measure the contribution of data owners
or participating clients. In multi-agent AI systems, they provide principled
reward allocation and contribution assessment \citep{kairouz2021advances,du2021survey}.

Fairness and stability do not always coincide. An allocation may reward
average marginal contributions fairly while still giving some coalition an
incentive to deviate. Convex games provide an important case where the two
goals align. A game is convex when
$v(S)+v(T)\leq v(S\cup T)+v(S\cap T)$ for all coalitions
$S,T\subseteq\mathcal{N}$. Convexity means that marginal contributions grow as
coalitions become larger, so cooperation exhibits increasing returns. In
convex games, the core is nonempty and the Shapley value lies in the core,
making the Shapley allocation both fair and coalitionally stable.

\subsection{Matching Games and Market Design}

Matching theory studies cooperation in markets where money may be absent or
insufficient to describe the allocation problem. Instead, agents have
preferences over potential partners, institutions, or assignments
\citep{gale1962college,maschler2013game}. Applications include school
admissions, labor markets, organ exchange, spectrum sharing, platform
assignment, and resource allocation in socio-technical systems.

In a two-sided matching market, let $M=\{m_1,\dots,m_n\}$ and
$W=\{w_1,\dots,w_n\}$ denote two sets of agents, and suppose each agent has a
strict preference ordering over agents on the opposite side. A matching
$\mu:M\rightarrow W$ assigns each agent in $M$ to exactly one agent in $W$.
The stability notion is based on blocking pairs: a pair $(m,w)$ blocks
$\mu$ if $m$ prefers $w$ to $\mu(m)$ and $w$ prefers $m$ to
$\mu^{-1}(w)$. A stable matching admits no such blocking pair.

The Gale-Shapley result shows that every finite two-sided matching market
with strict preferences admits at least one stable matching. The proof is
constructive through the Deferred Acceptance algorithm. In the
$M$-proposing version, unmatched agents in $M$ propose according to their
preference order, while agents in $W$ tentatively keep their most preferred
proposal and reject the rest. The process terminates in finitely many steps
and produces a stable matching that is optimal for all agents in $M$ among the
set of stable matchings.

Matching theory extends naturally to many-to-one and many-to-many settings,
including school choice, hospital-resident matching, task assignment, and
resource allocation with capacities. In such settings, stability often
corresponds to the absence of justified envy or blocking coalitions. The
central lesson for socio-technical systems is that cooperation is not only a
question of total value; it also depends on whether the assignment mechanism
respects incentives, preferences, and institutional constraints. Reputation and
matching mechanisms in networked socio-technical systems use similar ideas to
stabilize coalition formation and decentralized resource pairing
\citep{liu2011designreputationsystembased}. This
coalitional view reappears in Section~\ref{sec:tomohisa}, where incentives are
used to make desirable participation and coordination behavior sustainable.

\section{Feedback Learning and Control}\label{sec:control}

The game-theoretic models introduced in Sections~\ref{sec:tamer}
and~\ref{sec:quanyan} naturally lead to a feedback view of socio-technical
systems \citep{BZ18,bacsar1998dynamic,li2022confluence,zhu2012game,zhu2012differential}. Individuals,
organizations, and infrastructures repeatedly adapt their decisions in
response to observations, incentives, and environmental changes, connecting
game theory with control systems, distributed optimization, online learning,
adaptive decision-making, and reinforcement learning \citep{ZYB21}.

\begin{figure}[t]
    \centering
    \includegraphics[width=0.94\columnwidth]{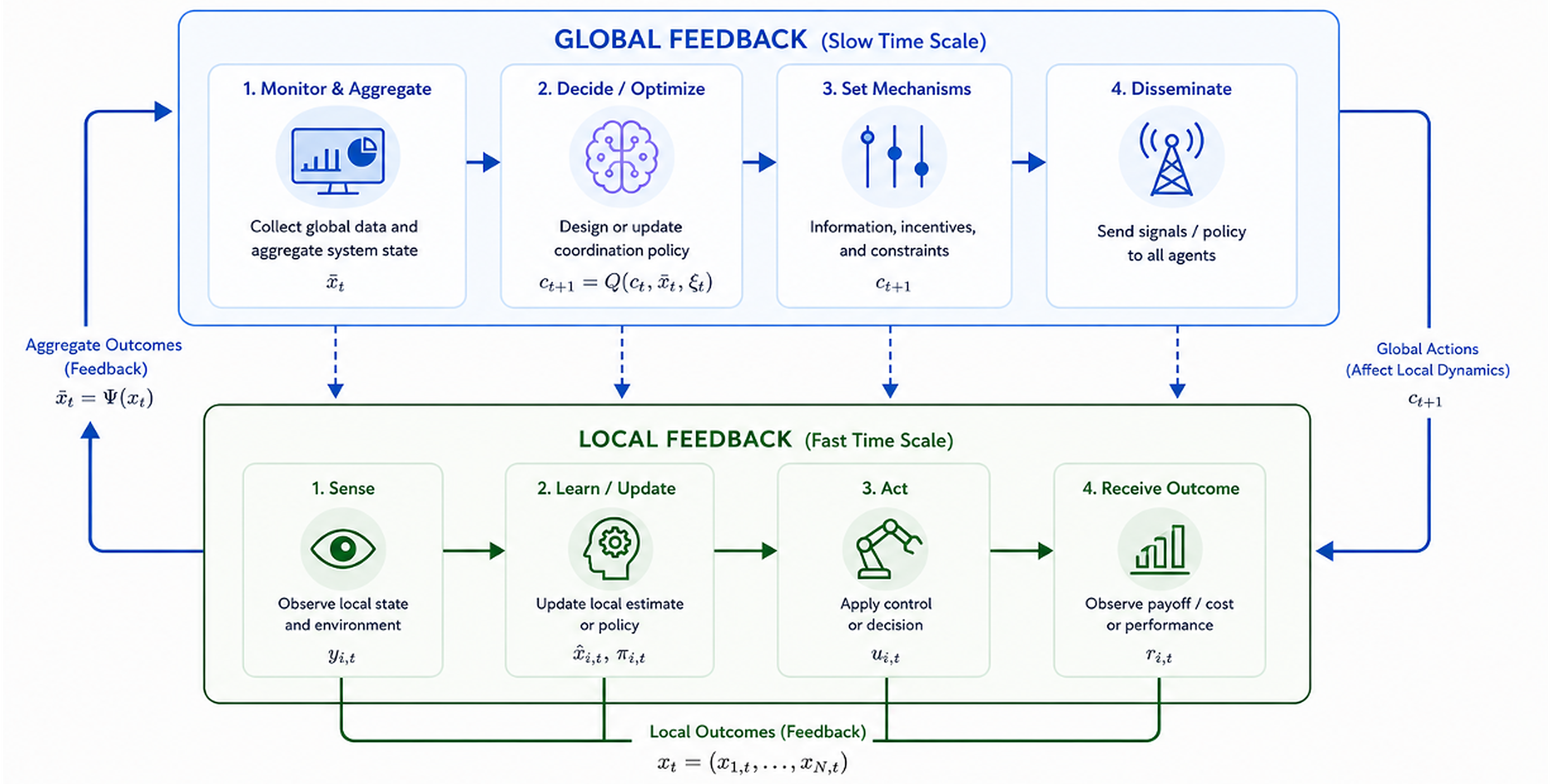}
    \caption{
    Coupled local and global feedback architecture in socio-technical systems.
    }
    \label{fig:global_local_feedback_architecture}
\end{figure}

A useful way to understand these dynamics is to separate the feedback architecture into two interconnected layers: \emph{local feedback learning} and \emph{global coordination and control}. The local layer captures decentralized adaptation among agents, whereas the global layer captures higher-level coordination and regulation. Figure~\ref{fig:global_local_feedback_architecture} illustrates this coupled local-global feedback structure. The lower layer represents fast decentralized feedback loops in which heterogeneous human, technical, and hybrid agents observe, learn, adapt, and interact with neighbors and the environment, generating aggregate behaviors and emergent collective patterns. The upper layer represents slower global coordination and control, where a coordinator aggregates system-wide information, updates coordination policies, and disseminates signals, incentives, and admissible constraints.

\subsection{Local Feedback Learning}

The first layer, \emph{local feedback learning}, describes how individual agents learn from their own experiences and from interactions with neighboring agents or the surrounding environment. Each agent observes local information, updates beliefs, and modifies behavior according to its own objectives and constraints. Because no single entity necessarily observes the full system, learning emerges through repeated local interactions.

Local agents are often heterogeneous. Human agents may be users, operators, organizations, decision-makers, or social communities, whereas technical agents may be autonomous systems, algorithms, robots, sensors, cyber infrastructures, software services, or machine-learning components. Hybrid human-machine agents combine these roles. Such agents are typically connected through communication, interaction, observation, and feedback networks.

Humans adapt to technological systems, recommendation engines, automated policies, and algorithmic decisions, while technical agents learn from human behaviors, preferences, responses, and operational patterns. These agents may learn in fundamentally different ways: human learning is often boundedly rational and context-dependent \citep{simon1990bounded,kahneman2013prospect}, whereas technical agents may use reinforcement learning, gradient optimization, Bayesian inference, adaptive filtering, evolutionary computation, or distributed optimization.

Using the notation of Section~\ref{sec:tamer}, let $x_{i,t}\in\mathcal{X}_i$ denote the strategy or action selected by agent $i$ at time $t$, and let $x_t=(x_{1,t},\dots,x_{N,t})$ denote the joint action profile. When a separate physical or informational state is needed, we write it as $z_t$, consistent with the dynamic-game notation introduced earlier.

At each stage, agents observe information from the environment and update their strategies according to a rule such as \(x_{i,t+1}=F_i(x_{i,t},I_{i,t})\), where $I_{i,t}$ denotes the information available to agent $i$ at time $t$. To capture heterogeneous adaptation mechanisms, one may generalize the update rule as \(x_{i,t+1}=F_i^{(\theta_i)}(x_{i,t},I_{i,t})\), where $\theta_i$ denotes the learning architecture, behavioral model, or adaptation type of agent $i$. For example, one subset of agents may follow RL dynamics \(x_{i,t+1}=F_i^{\mathrm{RL}}(x_{i,t},I_{i,t})\), another subset may employ best-response adaptation \(x_{j,t+1}\in BR_j(I_{j,t})\), and another may evolve according to evolutionary or imitation dynamics \(x_{k,t+1}=F_k^{\mathrm{EV}}(x_{k,t},I_{k,t})\). The information available to the agent may include neighboring actions, congestion levels, prices, historical observations, recommendations, or estimates of the system state, and is often decentralized, noisy, delayed, and heterogeneous.

A particularly important example is the \emph{best-response dynamic} \(x_{i,t+1}\in BR_i(I_{i,t})\) \citep{fudenberg1991}, where $BR_i(\cdot)$ denotes the best-response correspondence. Intuitively, the agent selects a strategy that maximizes its utility given its current information. In many settings, the best response is obtained from the optimization problem
\[
BR_i(I_{i,t})
=
\arg\max_{x_i\in\mathcal{X}_i}
J_i(x_i,x_{-i,t};c_t),
\]
where $J_i$ denotes the utility or payoff function of agent $i$, $x_{-i,t}$ denotes the strategies of the remaining agents, and $c_t$ is a coordination signal or environmental parameter. This formulation captures a key idea in game theory: agents make decisions not only according to the environment but also according to the anticipated behavior of other agents.

Learning plays a fundamental role in these local feedback dynamics because agents may lack accurate models of the environment, the objectives of other agents, or the long-term consequences of their actions. Depending on the application, agents may use reinforcement learning, Bayesian learning, no-regret learning, fictitious play, stochastic approximation, evolutionary adaptation, mirror-descent dynamics, gradient play, or online optimization to refine their strategies \citep{li2022confluence,ZYB21,zhang2021decentralizedmarl,sandholm2020evolutionary,zhu2010heterogeneous,li2024conjectural}.

A broad class of adaptive learning dynamics can be represented through coupled payoff-estimation and policy-update equations. Let $\hat{q}_{i,t}\in\mathbb{R}^{|\mathcal{X}_i|}$ denote the estimated payoff vector of agent $i$ over its finite action set $\mathcal{X}_i$, and let $\pi_{i,t}\in\Delta(\mathcal{X}_i)$ denote the mixed strategy or policy used by the agent at time $t$. A generic payoff-learning update can be written as
\[
\hat{q}_{i,t+1}
=
(1-\mu_{i,t})\hat{q}_{i,t}
+
\mu_{i,t}
G_i\!\left(
\pi_{i,t},
\hat{q}_{i,t},
I_{i,t},
x_{i,t}
\right),
\]
where $G_i(\cdot)$ represents the payoff-learning operator and $\mu_{i,t}$ is the learning rate. This update captures how agents iteratively refine local payoff estimates using observations gathered from repeated interactions with the environment and neighboring agents.

Based on updated payoff estimates, the agent subsequently adjusts its strategy according to a policy adaptation rule of the form
\[
\pi_{i,t+1}
=
(1-\lambda_{i,t})\pi_{i,t}
+
\lambda_{i,t}
\Psi_i\!\left(
\pi_{i,t},
\hat{q}_{i,t+1},
I_{i,t},
x_{i,t}
\right),
\]
where $\Psi_i(\cdot)$ denotes the policy-update operator and $\lambda_{i,t}$ is the policy adaptation step size. Intuitively, actions associated with higher estimated payoffs are reinforced and therefore selected with greater probability in future interactions.

This coupled learning structure provides a unifying representation for adaptive schemes arising in game theory, reinforcement learning, stochastic approximation, and distributed control. Different choices of $G_i$ and $\Psi_i$ recover reinforcement learning, actor-critic learning, fictitious play, smooth best-response dynamics, replicator dynamics, no-regret learning, and Bayesian adaptation \citep{gao2021passivity}. When the reward or transition model is uncertain, robust and risk-sensitive variants of stochastic games and multi-agent reinforcement learning provide complementary ways to guard against model misspecification and adverse environments \citep{zhang2020robustmarl,TB21,mao2023provably}.

Heterogeneous learning agents can generate nontrivial collective behaviors. Technical agents may amplify or suppress human coordination patterns through recommendation systems, pricing algorithms, automated moderation, or adaptive control policies. Conversely, human responses may reshape the learning environment of technical agents by altering data distributions, operational contexts, and observed reward structures. Such mutual adaptation can lead to synchronization, polarization, cascading failures, correlated coordination, feedback amplification, or instability.

An important objective in studying such dynamics is to understand whether repeated interactions converge to meaningful long-term behaviors. Under suitable assumptions, the resulting system may converge to equilibrium concepts such as Nash equilibria, correlated equilibria, coarse correlated equilibria, quantal response equilibria, or mean-field equilibria \citep{nash1950,caines2021mean,tembine2013risk}. It is important, however, to distinguish between an equilibrium concept and the learning process that may generate it. A Nash equilibrium is a static solution concept, whereas learning dynamics describe how agents update strategies over time. When a game has multiple equilibria, the learning rule, information structure, initialization, noise process, and network topology may determine which equilibrium is selected and whether it is dynamically stable.

From a dynamical system perspective, the equilibrium points of the induced learning dynamics correspond to stationary points or invariant sets of the coupled feedback system. In many settings, Nash equilibria arise as fixed points of best-response, policy-iteration, or gradient-based adaptation dynamics. Stability analysis of these dynamical systems, therefore, becomes closely connected to equilibrium analysis in game theory. Concepts such as Lyapunov stability, attractors, invariant manifolds, and variational stability provide analytical tools for understanding whether strategic learning converges, oscillates, diverges, or exhibits complex emergent behavior \citep{frihauf2012nash,oliveira2021nash,poveda2022fixed}.

Historically, these dynamics were primarily studied for equilibrium computation and equilibrium selection. Today, they are also used for distributed control, adaptive coordination, online optimization, resilient autonomy, and decentralized intelligence in large-scale socio-technical systems. The incentive mechanisms in Section~\ref{sec:tomohisa} can be viewed as one way of shaping these feedback dynamics.

\subsection{Global Coordination and Control}

An important feature of socio-technical systems is the presence of a higher-level designer or regulator that seeks to shape collective system behavior. While agents adapt locally according to their own objectives, a centralized coordinator may influence the overall system dynamics toward broader societal or operational goals such as efficiency, fairness, resilience, sustainability, security, or social welfare.

As illustrated in Fig.~\ref{fig:global_local_feedback_architecture}, the global coordination layer often operates on a slower time scale than the decentralized local feedback loops. The coordinator aggregates system-wide observations, updates coordination policies, and disseminates information, incentives, and operational constraints back to the agents.

The coordinator introduces a coordination signal $c_t$ to guide the system toward desired objectives. The information available to agents may depend on this intervention: \(I_{i,t}=\Phi_i(z_t,x_t,c_t,\omega_t)\), where $\Phi_i(\cdot)$ represents the information-generation mechanism and $\omega_t$ represents uncertainty or external disturbances.

The coordination signal may correspond to pricing mechanisms, incentives, routing recommendations, communication policies, or resource allocations. Global coordination is often only partially observable and partially controllable. A coordinator may influence only a subset of agents, infrastructures, organizations, or communication channels, so coordination signals propagate indirectly through local interactions.

Let \(\mathcal{C}_t \subseteq \mathcal{N}\) denote the subset of agents directly influenced by the coordinator at time $t$. The controlled agents evolve according to \(x_{i,t+1}=F_i(x_{i,t},I_{i,t},c_t)\) for \(i\in\mathcal{C}_t\), whereas the remaining agents adapt primarily through decentralized interactions according to \(x_{j,t+1}=F_j(x_{j,t},I_{j,t})\) for \(j\notin\mathcal{C}_t\). Even limited control over a subset of nodes may reshape aggregate behavior through cascades of local interactions, adaptation, and network effects.

One important mechanism for global coordination is \emph{information control} \citep{bergemann2019information,akyol2016information,massicot2019public,velicheti2023strategic,zhang2021informational,zhang2023stochastic,simhon2016optimalinformationdisclosurepoli}. Rather than directly prescribing actions, the coordinator influences the information available to the agents through public or private signals. These may include alerts, forecasts, reputation scores, recommendations, risk assessments, or partial state observations. The information structure may therefore be represented as \(I_{i,t}=\Phi_i(z_t,x_t,\sigma_t^{\mathrm{pub}},\sigma_{i,t}^{\mathrm{priv}},\omega_t)\), where $\sigma_t^{\mathrm{pub}}$ denotes a public signal observable by all agents and $\sigma_{i,t}^{\mathrm{priv}}$ denotes a private signal provided specifically to agent $i$.

Through these information channels, the coordinator shapes beliefs, expectations, and strategic responses without explicitly enforcing actions. Another important coordination mechanism is \emph{incentive design} \citep{myerson1991,basar2024inducement,zhu2012guidex,zhang2022incentive,chen2021dynamic}. In this setting, the coordinator modifies the reward structure perceived by the agents in order to induce desirable collective behavior. Let \(\widetilde J_{i,t}=J_i(z_t,x_{i,t},x_{-i,t};c_t)+R_i(z_t,x_t,c_t)\) denote the modified utility function, where $R_i(\cdot)$ represents incentives or penalties introduced by the coordinator. These modifications may correspond to taxes, subsidies, reputation adjustments, congestion pricing, or energy-saving incentives designed to align local objectives with system-level goals.

Under incentive-based coordination, the local adaptation dynamics become \(x_{i,t+1}=F_i(x_{i,t},I_{i,t};\widetilde J_{i,t})\), showing that the learning dynamics themselves depend on the modified reward structure induced by the global controller.

Beyond modifying information and rewards, the coordinator may also alter the \emph{feasible action space} available to the agents. Let \(\mathcal{X}_{i,t}(c_t)\subseteq\mathcal{X}_i\) denote the admissible strategy set of agent $i$ under coordination policy $c_t$. The coordinator may therefore impose operational constraints, security restrictions, access controls, routing limitations, or resource constraints that dynamically reshape the feasible decision space. The constrained update selects \(x_{i,t+1}\) according to \(F_i(x_{i,t},I_{i,t})\) while enforcing \(x_{i,t+1}\in\mathcal{X}_{i,t}(c_t)\). Such admissible-set modifications arise naturally in cybersecurity, transportation, smart grids, and financial systems, where policies may restrict access, routes, power usage, or transactions.

The coordination policy itself may evolve adaptively according to observed system behavior, \(c_{t+1}=Q(H_t,c_t)\), where $H_t$ denotes the history of system observations available to the coordinator and $Q(\cdot)$ describes the coordination update mechanism.

\subsection{Interdependence between Global and Local Dynamics}

The local and global dynamics in socio-technical systems are intertwined, and their interactions often determine long-term system behavior. This interaction provides the bridge from the feedback architecture in Section~\ref{sec:control} to the incentive and resilience questions in Sections~\ref{sec:tomohisa} and~\ref{sec:hideaki}.

Decentralized local interactions can generate emergent global behaviors without explicit centralized coordination. Through repeated adaptation, heterogeneous agents may produce large-scale organization, conventions, correlated behavior, and equilibrium structures in markets, social networks, transportation systems, and distributed cyber-physical infrastructures. Conversely, the global coordination layer directly influences local learning by modifying information structures, reward structures, and admissible actions, and aggregate outcomes then feed back into the global coordination mechanism.

An important characteristic of socio-technical systems is the separation of time scales between local adaptation and global coordination. Local learning processes are typically fast, whereas global coordination often requires system-wide information aggregation, policy computation, organizational decision-making, and dissemination of control signals. This separation naturally leads to a \emph{two-time-scale dynamical system}.

The coexistence of fast local adaptation and slow global coordination creates important challenges for stability, robustness, and control design. A centralized coordination policy that neglects local feedback effects may alter incentives or constraints in ways that destabilize the existing collective order. A classical example is the Braess paradox in transportation networks, where adding infrastructure or routing flexibility can worsen performance because strategic local adaptations alter the equilibrium traffic distribution \citep{braess1968}. Similar effects arise in communication networks, financial systems, cybersecurity, and distributed resource-allocation systems, where well-intentioned global interventions may amplify congestion, fragility, instability, or adversarial behavior. Consequently, effective socio-technical control requires coordination mechanisms that preserve, stabilize, and exploit desirable emergent properties arising from decentralized interactions.

From a control-theoretic perspective, socio-technical systems are hierarchical multi-time-scale feedback systems involving local learning, emergent collective behavior, and slower global coordination. Their analysis requires ideas from game theory, adaptive control, distributed optimization, dynamical systems, learning theory, and network science.

The same perspective extends to cooperative games and coalition formation \citep{maschler2013game,jackson2002coalition}. Let $\mathcal{P}_t=\{S_1,\dots,S_m\}$ denote the coalition structure formed at time $t$, where each coalition satisfies $S_k\subseteq\mathcal{N}$. Agents interact locally while adapting to coalition-level signals and global coordination mechanisms. Global incentives, signaling policies, or resource-allocation rules may induce merge-and-split dynamics, exit-and-join dynamics, and adaptive bargaining, while local interactions within coalitions influence coalition stability, cooperation incentives, and collective performance.

Following the cooperative-game notation of Section~\ref{sec:quanyan}, $v(S)$ denotes the value generated by coalition $S\subseteq\mathcal{N}$. The utility obtained by agent $i$ may depend on both local actions and coalition membership, represented by $J_i(x_t,S_{i,t};c_t)$, where $S_{i,t}$ denotes the coalition containing agent $i$ and $c_t$ denotes the global coordination signal. Coalition adaptation can evolve jointly with strategic dynamics according to $\mathcal{P}_{t+1}=\Gamma(\mathcal{P}_t,x_t,c_t)$, where $\Gamma(\cdot)$ is induced by local interactions, negotiation outcomes, incentives, and global coordination policies.

The global coordination layer may also regulate coalition sizes, resource distributions, or admissible coalition structures to achieve resilience, efficiency, fairness, or robustness. Similar principles arise in matching and resource-allocation systems, where agents adapt preferences and beliefs while the system designer modifies priorities, resources, or operational constraints.

\section{Incentive Mechanism Design}\label{sec:tomohisa}

The feedback-learning framework developed in Section~\ref{sec:control} naturally
leads to the problem of incentive design. In socio-technical systems, agents
adapt according to their own objectives, but a system-level designer often
cares about broader outcomes such as efficiency, fairness, sustainability,
resilience, and social welfare. Incentives provide an indirect control
mechanism: rather than prescribing every action, a coordinator modifies the
strategic environment so that decentralized responses move the system toward
desirable collective behavior \citep{myerson1991,basar1984affine,basar2024inducement}.

\subsection{Incentives as Feedback Signals}

Using the agent set $\mathcal{N}$ from Section~\ref{sec:tamer}, let
$x_{i,t}\in\mathcal{X}_i$ denote the action of agent $i$ at time $t$ and
let $x_t=(x_{1,t},\dots,x_{N,t})$ denote the joint action profile. As in
Section~\ref{sec:control}, $z_t$ denotes the system state when it is useful to distinguish the
environment from the agents' decisions. Agent $i$ has nominal payoff
$J_i(z_t,x_{i,t},x_{-i,t};c_t)$, where $c_t$ is a coordination signal or
environmental parameter.

An incentive mechanism modifies this payoff through an additional term
$\rho_{i,t}$, yielding
\[
\widetilde J_i(z_t,x_t;c_t,\rho_t)
=
J_i(z_t,x_{i,t},x_{-i,t};c_t)+\rho_{i,t}(z_t,x_t,c_t).
\]
The incentive $\rho_{i,t}$ may represent a subsidy, tax, reward, penalty,
priority rule, access right, reputation adjustment, or resource-allocation
credit. The modified payoff changes both the equilibrium selected by the
agents and the transient learning dynamics that lead toward that equilibrium. If the local adaptation rule is
\(x_{i,t+1}=F_i(x_{i,t},I_{i,t};J_i)\), then incentive-based coordination
changes it to \(x_{i,t+1}=F_i(x_{i,t},I_{i,t};\widetilde J_i)\). Thus,
incentives act as feedback signals embedded in the agents' objective
functions. This is why incentive design belongs naturally with feedback
control: the designer shapes the closed-loop behavior of a strategic system
through the payoffs perceived by its participants.

\subsection{Pareto Improvement and Budget Constraints}

Incentive mechanisms are meaningful only when they respect participation and
resource constraints. A designer may seek a Pareto improvement, meaning that
the induced outcome improves at least one agent's payoff without reducing any
other agent's payoff relative to a baseline equilibrium or trajectory. At the
same time, incentives must be sustainable: the total subsidies, taxes, or
transfers cannot require an unlimited external budget.

Let $\bar x_t$ denote the baseline behavior generated without the new
incentive, and let $x_t^\rho$ denote the behavior induced by incentive policy
$\rho_t$. A Pareto-improving incentive policy can be expressed informally as
one satisfying
\[
J_i(z_t,x_t^\rho;c_t)+\rho_{i,t}
\geq
J_i(z_t,\bar x_t;c_t),
\qquad i\in\mathcal{N},
\]
with strict improvement for at least one agent over the relevant time horizon.
A sustainable budget constraint requires the aggregate incentive expenditure
to remain bounded, for example through a condition such as
\(\sum_{t\geq0}\delta^t\sum_{i\in\mathcal{N}}\rho_{i,t}\leq B\), where
$B$ is the available budget and $\delta\in(0,1]$ discounts future transfers.

This formulation captures a central design tradeoff. Strong incentives may
quickly move agents toward a desirable behavior, but they can violate budget
limits or create dependency on external subsidies. Weak incentives may be
budget-feasible but fail to change strategic behavior. Pareto-improving
mechanisms under sustainable budget constraints address this tradeoff by
seeking incentive rules that improve welfare without imposing a persistent
unfunded cost \citep{yan2023pareto}.

\subsection{Hierarchical Incentive Structures}

Many socio-technical systems contain multiple organizational layers. Agents
may belong to groups, platforms, firms, regions, fleets, or communities, and
these groups may themselves interact strategically. Partition \(\mathcal{N}\) as
\(\mathcal{N}_1\cup\cdots\cup\mathcal{N}_M\), where the subsets represent
$M$ groups. An intragroup incentive acts within a group
$\mathcal{N}_m$, while an intergroup incentive acts across groups by shaping
how group-level objectives interact.

For agent $i\in\mathcal{N}_m$, a hierarchical incentive can be written as
\[
\widetilde J_i
=
J_i(z_t,x_{i,t},x_{-i,t};c_t)
+\rho^{\mathrm{in}}_{i,t}
+\rho^{\mathrm{out}}_{m,t},
\]
where $\rho^{\mathrm{in}}_{i,t}$ is an intragroup incentive assigned to agent
$i$, and $\rho^{\mathrm{out}}_{m,t}$ is an intergroup incentive associated
with group $m$. The intragroup term can promote coordination among members of
the same group, while the intergroup term can align the behavior of different
groups with system-level objectives. This hierarchy is useful in transportation, energy, communication, and
organizational systems. For example, a fleet operator may incentivize vehicles
within a fleet to balance load locally, while a city authority uses
intergroup incentives to coordinate multiple fleets. Similarly, a power-system
operator may shape the behavior of users within each region while also
coordinating demand across regions. Hierarchical noncooperative dynamical
systems formalize these layered interactions and show how intragroup and
intergroup incentives can reshape equilibrium behavior and closed-loop
dynamics \citep{yan2023hierarchical}.

\subsection{Design Implications}

Incentive design links the noncooperative, cooperative, and control
perspectives developed in Sections~\ref{sec:tamer}--\ref{sec:control}. From
the noncooperative viewpoint, incentives change individual payoffs and
equilibrium responses; from the cooperative viewpoint, they can support
participation, sharing, and fair allocation by making cooperation individually
attractive; and from the control viewpoint, they are feedback inputs that shape
learning dynamics under information and budget constraints. The main design challenge is to choose incentives that are effective,
credible, and sustainable. Effective incentives change behavior in the desired
direction. Credible incentives respect the information and authority available
to the coordinator. Sustainable incentives obey budget, fairness, and
participation constraints over time. Poorly designed mechanisms may instead
amplify instability, induce manipulation, exhaust budgets, or shift risk to
other parts of the system. These issues connect directly to the resilience and
security questions in Section~\ref{sec:hideaki}.

\section{Resilience and Security in Multi-Agent Systems}\label{sec:hideaki}

The feedback-learning and incentive-design frameworks in
Sections~\ref{sec:control} and~\ref{sec:tomohisa} provide a foundation for
studying resilience and security in socio-technical multi-agent systems
\citep{zhu2015game,huang2020dynamic,ishii2022security,zhu2024foundations,zhu2024disentangling,zhu2019multilayer}. In realistic environments,
agents operate under uncertainty, failures, misinformation, cyberattacks,
adversarial manipulation, and infrastructure disruptions.

\begin{figure}[t]
    \centering
    \includegraphics[width=0.94\columnwidth]{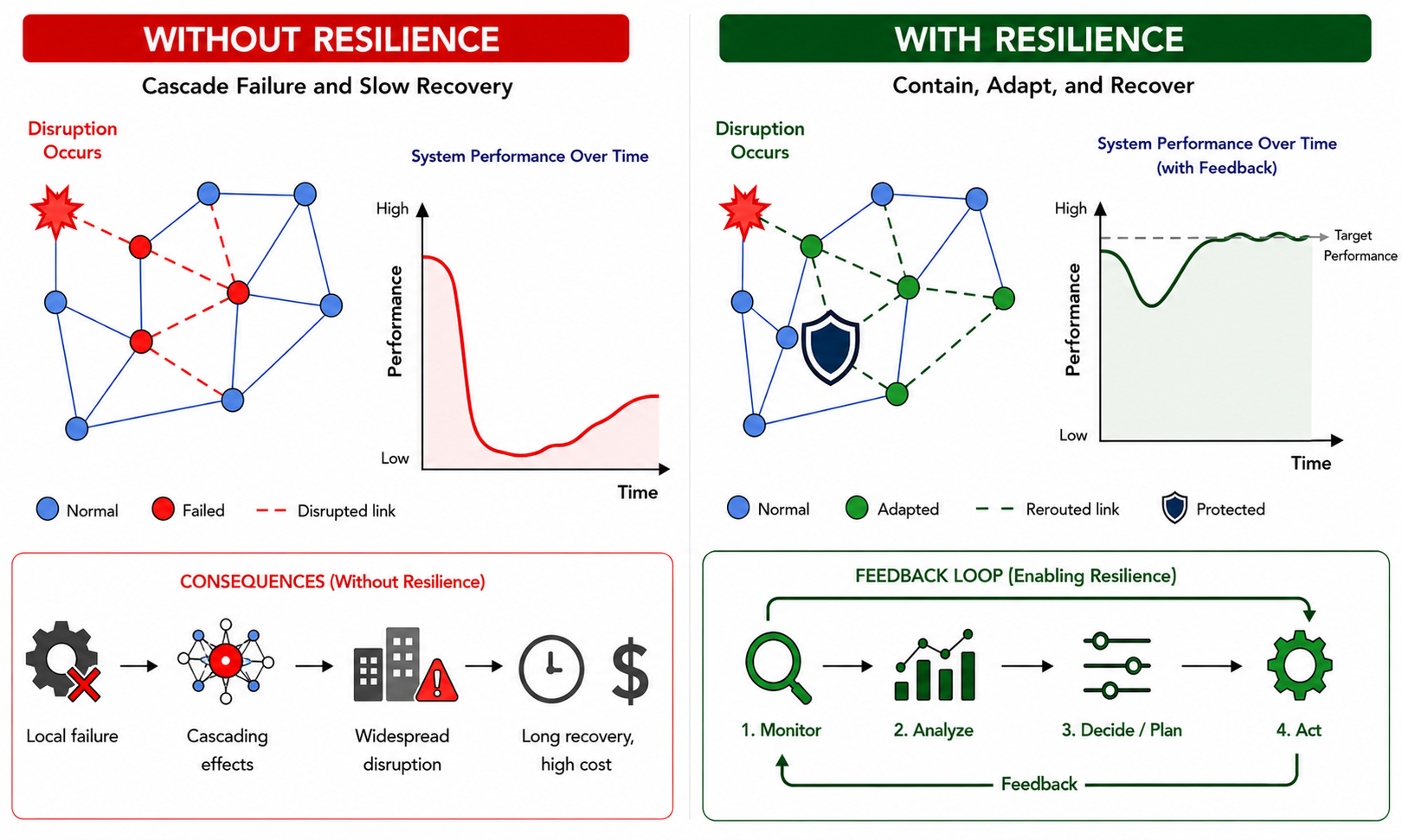}
    \caption{
    Comparison of socio-technical systems without and with resilience.
    Without resilience, disruptions propagate and cause cascading failures with slow recovery.
    With resilience, feedback-based adaptation enables monitoring, coordination, containment, and rapid recovery.
    }
    \label{fig:resilience_feedback}
\end{figure}

As illustrated in Fig.~\ref{fig:global_local_feedback_architecture}, vulnerabilities may arise at multiple levels of the coupled local-global feedback architecture. At the local level, individual agents may be compromised, manipulated, or deceived. At the network level, communication channels and information flows may be corrupted or disrupted. At the global level, coordination policies, incentive mechanisms, and supervisory information structures may themselves become attack surfaces \citep{zhu2020cross}. As illustrated in Fig.~\ref{fig:resilience_feedback}, resilience emerges through a feedback loop that enables monitoring, adaptation, and coordinated recovery under disruptions.

A fundamental challenge is that adversaries are adaptive. Attackers learn from defensive responses, exploit structural vulnerabilities, and revise strategies over time, causing the local learning dynamics to evolve under both nominal and adversarial influences \citep{manshaei2013game,pawlick2019game}. Let \(z_t\) denote the system state and let \(a_t\) denote adversarial actions.
The resulting dynamics can be represented abstractly as
\(
z_{t+1}=F(z_t,a_t),
\)
where \(a_t\) may include misinformation injection, spoofing, communication corruption, or strategic deception.

In many socio-technical systems, adversaries primarily target the information structure itself. Let \(I_{i,t}\) denote the information available to agent \(i\) at time \(t\), consistent with Section~\ref{sec:control}. Under adversarial influence, agent \(i\) may instead observe corrupted information
\(
\widetilde I_{i,t}
=
\Gamma_{i,t}\!\left(
I_{i,t},
a_t
\right),
\) 
where \(\Gamma_{i,t}(\cdot)\) represents an adversarial transformation induced by misinformation injection, filtering, spoofing, or communication corruption \citep{yang2023designing,amini2024control}. Since the decentralized feedback loops in Fig.~\ref{fig:global_local_feedback_architecture} are tightly interconnected, localized attacks may propagate through the interaction network and generate systemic disruptions. Dynamic-game models of infrastructure protection and recovery make this coupling explicit by linking attacker-defender decisions, network interdependence, and restoration policies \citep{chen2017infrastructure,chen2019dynamic,chen2021secureiot,chen2016interdependent,huang2017large,huang2018factored}.

Resilience therefore requires maintaining stability, adaptability, and coordination despite disturbances or attacks. At the local level, agents may employ resilient learning mechanisms such as filtering, trust estimation, anomaly detection, and robust aggregation \citep{zhao2022multi,huang2022reinforcement}. Dynamic resilient network games provide one way to model how agents preserve consensus, connectivity, and cluster structure under jamming or disrupted communication \citep{nugraha2020dynamic,nugraha2020jamming,nugraha2020connectivity,nugraha2021rolling}. For example, agents may adapt trust weights \(w_{ij,t}\) associated with neighboring agents in order to reduce the influence of unreliable or compromised information sources.

The global coordination layer can further enhance resilience through information control, incentive design, and operational constraints. Trusted public signals, reputation mechanisms, and defensive incentives may encourage secure and cooperative behavior while limiting adversarial propagation. Operational constraints may additionally isolate compromised agents, restrict vulnerable communication pathways, or dynamically reshape feasible interaction structures during attack conditions.

Because socio-technical systems are highly interconnected, local disruptions may trigger cascading failures through communication networks, infrastructure dependencies, or correlated strategic responses. Human agents and AI systems also respond differently to incentives, uncertainty, and misinformation, creating vulnerabilities at the interface between social and technical subsystems.

Resilient socio-technical systems therefore require adaptive and self-healing capabilities. Local agents must continuously update trust relationships and interaction patterns, while the global coordination layer adjusts incentives, signaling policies, and operational constraints in response to evolving threats \citep{nugraha2022rolling}. From this perspective, resilience emerges from the coupled multi-time-scale interaction of decentralized learning, network adaptation, and global coordination, tying the security discussion back to the full modeling arc developed in Sections~\ref{sec:tamer}--\ref{sec:tomohisa}.

\section{Concluding Remarks and Future Directions}\label{sec:conclusion}

 This tutorial has introduced game-theoretic methods for socio-technical
systems, moving from noncooperative and cooperative foundations in
Sections~\ref{sec:tamer} and~\ref{sec:quanyan} to feedback learning,
incentive design, and resilience in Sections~\ref{sec:control}--\ref{sec:hideaki}.
These frameworks provide complementary tools for modeling uncertainty,
large-population interactions, strategic adaptation, and mechanism design.

Because socio-technical systems are composed of many interacting agents, infrastructures, and organizations, effective modeling often requires a bottom-up perspective that captures decentralized interactions and emergent collective behavior. Their evolving attack surfaces also create persistent challenges involving threats, vulnerabilities, systemic risks, and adversarial behavior.

For clarity, this tutorial has discussed cooperative and noncooperative paradigms separately. In practice, however, these paradigms frequently coexist and interact within the same socio-technical system. This motivates the development of unified hybrid frameworks capable of simultaneously capturing competitive and cooperative behaviors. Models that integrate cooperation and competition, including cooperative-noncooperative and hierarchical game-theoretic formulations, together with their associated learning, adaptation, and control challenges, constitute an important direction for future research.

Control theory, particularly distributed and adaptive control, remains central to this agenda. Future work must better connect equilibrium concepts with feedback mechanisms, multi-time-scale adaptation, closed-loop learning, resilience, robustness, and the propagation of uncertainty.

\section*{Acknowledgements}
This work was supported in part by CNS-2521130 from NSF, and in part by FA9550-24-1-0152 from AFOSR.

%\section*{DECLARATION OF GENERATIVE AI AND AI-ASSISTED TECHNOLOGIES IN THE WRITING PROCESS}

%During the preparation of this work, the author used ChatGPT (OpenAI) to assist with language refinement, organization of LaTeX sections, and clarification of expository passages.  All mathematical formulations, theoretical results, algorithms, simulations, and scientific interpretations were conceived, developed, and validated by the author. After using the tool, the author thoroughly reviewed and edited all content as needed and takes full responsibility for the accuracy and integrity of the final manuscript.

\bibliographystyle{ifacconf}
\bibliography{ifacconf,quanyan_zhu_google_scholar}

\end{document}